\documentclass[manuscript]{aastex}
\shorttitle{Three-minute umbral oscillations}
\shortauthors{Reznikova et al.}

\begin{document}

\title{Three-minute oscillations above sunspot umbra observed with SDO/AIA and NoRH}
\author{V. E. Reznikova\altaffilmark{1}, 
K. Shibasaki\altaffilmark{1}, 
R.~A. Sych\altaffilmark{2,3},
V.~M. Nakariakov\altaffilmark{4,5} 
\and }

\altaffiltext{1}{Nobeyama Solar Radio Observatory/NAOJ, Nagano 384-1305, Japan; reznik@nro.nao.ac.jp}
\altaffiltext{2}{Key Laboratory of Solar Activity, National Astronomical Observatories, Beijing 100012, China}
\altaffiltext{3}{Institute of Solar-Terrestrial Physics, P.O.Box 4026, Irkutsk, Russia}
\altaffiltext{4}{Physics Department, University of Warwick, Coventry, CV4 7AL, UK}
\altaffiltext{5}{Central Astronomical Observatory at Pulkovo of the Russian Academy of Sciences, 196140 St Petersburg, Russia}

\begin{abstract}
Three-minute oscillations over sunspot's umbra in AR 11131 were observed simultaneously in UV/EUV emission by SDO/AIA and in radio emission by Nobeyama Radioheliograph (NoRH). We use 24-hours series of SDO and 8-hours series of NoRH observations to study spectral, spatial and temporal variations of pulsations in the 5-9 mHz frequency range at different layers of the solar atmosphere. High spatial and temporal resolution of SDO/AIA in combination with long-duration observations allowed us to trace the variations of the cut-off frequency and spectrum of oscillations across the umbra. We found that higher frequency oscillations are more pronounced closer to the umbra's center, while the lower frequencies concentrate to the peripheral parts. We interpreted this discovery as a manifestation of variation of the magnetic field inclination across the umbra at the level of temperature-minimum. Possible implications of this interpretation for the diagnostics of sunspot atmospheres is discussed.
\end{abstract}

 
\section{Introduction}
Despite the fact that three-minute sunspot oscillations have been observed for many decades and enormous theoretical progress has been made in understanding this phenomenon \citep[for reviews see][]{Bogdan, Thomas, Khomenko}, there are still many questions concerning sunspot wave physics. It is generally accepted, that they are slow magnetoacoustic waves moving with the local sound speed along magnetic field lines from the photosphere through the chromosphere into the corona. However, it is still unclear, how high in the corona they can propagate? What determines their properties in the corona? What is the spatial variation of the cut-off frequency across the umbra?   Are they directly linked with the propagating longitudinal waves ubiquitously observed in the EUV fans of coronal active regions \citep[e.g. see][for recent reviews]{De Moortel, Verwichte}. What is the nature of the recently revealed relationship of the three-min oscillations and similar periodicities in flaring energy releases \citep{Sych}? Understanding these issues provides an opportunity for plasma diagnostic at different layers of solar atmosphere since these waves convey the information about the properties of their formation and propagation mediums. 
Thus, \cite{Shibasaki} proposed that umbral oscillations seen in microwaves can be used as a diagnostic of the temperature-minimum region. Further \cite{Solanki} pointed out that observations with higher spatial resolution are needed to isolate the umbra and to improve the diagnostics. 

Spatial distribution of 3-min oscillation frequency and power across a sunspot at different heights has been intensively studied with the use of data obtained with ground-based and space-borne instruments (e.g. onboard SOHO, TRACE and Hinode). In particular, SOHO/MDI observations showed that at the photosphere the oscillation power is enhanced in the Doppler velocity and line-depth signals, but not in the continuum intensity \citep{Jain-2002}. Enhancement of the spectral power was found in the regions of reasonably strong magnetic field, but showed evident suppression in regions of strongest fields, in the center of the sunspot. Observations in $H_\alpha$ demonstrated that standing oscillations with the frequencies around 6.5 mHz were dominant in the umbral and inner penumbral regions, together with the presence of running penumbral waves in the band of 3 mHz. However these waves were not detected at the photospheric heights \citep{Christopoulou-2000, Christopoulou-2001}.  It was suggested that both 3-min umbral oscillations and 5-min running penumbral waves were in fact different manifestations of the same wave, generated in the photosphere, and propagating along different magnetic field lines with the different inclination angle \citep{van der Voort, Bloomfield}. \cite{Christopoulou-2003} demonstrated that oscillations with different frequencies in the 3-min band were localized within different parts of the umbral chromosphere. Then \cite{Tziotziou-2006} found that in the chromosphere, the frequency of 3-min oscillations decreases from the umbra outwards, with the abrupt change at the umbra-penumbra boundary. A small \lq\lq node'' at the center of the umbra, where the oscillations with the frequencies above 4 mHz were suppressed, was reported by \cite{Nagashima-2007}. A similar effect of \lq\lq calmest umbral position'' was found in the Doppler velocity maps in Ca II 8542 \AA\ by \cite{Tziotziou-2007}. Nature of this \lq\lq black spot'' in the spatial distribution of the chromospheric 3-min oscillations remains unrevealed.

In the transition region the 3-min oscillations are reported to fill in all the sunspot umbra. In the corona they concentrate in smaller regions that coincide with the endpoints of bright coronal loops \citep{Brynildsen-2000,Brynildsen-2002,Brynildsen}. \cite{De Moortel-2002} used TRACE observations of EUV intensity waves in 38 coronal loops and found that in the loops situated above sunspots the waves have periods of about 3 minutes. This suggests that the underlying oscillations can propagate through the transition region into the corona.

Hence, it is still not clear how the oscillation power and frequency in the 3-min band are distributed within the umbra of the sunspot at different altitudes. This knowledge is important for revealing the nature of 3-min oscillations, and for diagnostic of the solar atmosphere above sunspots.

The data from the Atmospheric Imaging Assembly \citep[AIA;][]{Title} on the Solar Dynamics Observatory \citep[SDO;][]{Schwer}, with their high resolution (0.6 arcsec per pixels) and high cadence (up to 12 s), offer a new opportunity for such a study.  The instrument observes solar plasma from photospheric to coronal temperatures, taking full-disk images in a variety of extreme ultraviolet (EUV), ultraviolet (UV) and visible light wavelength bands. The other advantage of SDO data is 24-hour continuous observations. Three-min oscillations have very high quality, they can be observed for many days. Therefore, long-duration observations allow us to study this phenomenon with very high spectral resolution, which has not been done before.

In this paper we analyze simultaneous observations of the sunspot-associated oscillations by two instruments, AIA and the Nobeyama Radioheliograph, in order to obtain the information on spectral, spatial and temporal structures of oscillations above the same sunspot. We restrict our analysis to the frequency range 5-9 mHz and horizontal extent to the umbral area. 

\section{Observations}\label{Obs}
\begin{figure}
\begin{center}
\includegraphics[totalheight=17cm]{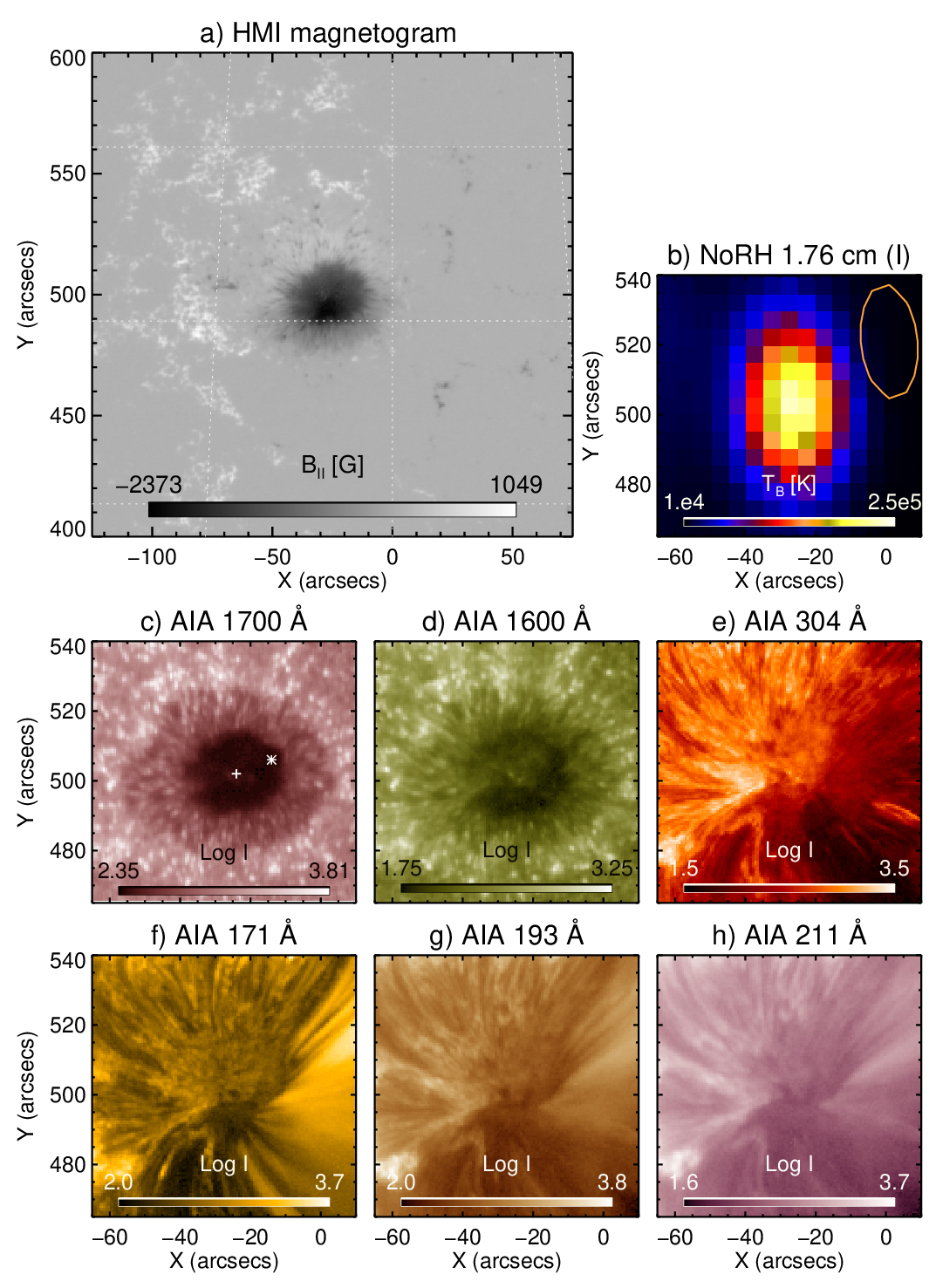}
\end{center}
\caption{(a) SDO/HMI magnetogram of the sunspot region in NOAA AR 11131 and images of sunspot taken (b) by NoRH and (c-h) by SDO/AIA on 2010 December, 8 at 04:00 UT. Orange ellipse at top-right of plot (b) indicates the half-maximum NoRH beam size. White cross and asterisk on plot (c) indicate two pixels inside the umbra for later reference.}
\label{fig1}
\end{figure}

Active region (AR) 11131 was available for observations in the northern hemisphere near to 30 degrees latitude  
and crossed the central meridian 2010 December 8. We selected this date for our analysis. Figure \ref{fig1}(a) presents line-of-sight magnetogram of the sunspot region from Helioseismic and Magnetic Imager (HMI) on SDO.
AR consisted of a well-developed sunspot of southern polarity with the magnetic field strong enough, $B=-2300$ G to produce significant and highly polarized emission at 17 GHz ($\lambda=1.76$ cm). The sunspot was surrounded by a number of small magnetic fragments of northern polarity. 

The emission mechanism of radio sources associated with sunspots is known to be gyroresonance
emission \citep[e.g.][]{Zheleznyakov}. At 17 GHz, the third harmonic layer is the most probable emission region, corresponding to a 2000 G isogauss layer \citep{Shibasaki94}.

The radio map of the sunspot that we studied obtained by Nobeyama Radioheliograph (NoRH) is shown in Fig. \ref{fig1}(b). Color indicates brightness temperature at 1.76 cm (Stokes I). NoRH measures both right circular polarization and left circular polarization. Using these measurements we synthesized a set of NoRH images in I and V Stokes parameters with time interval of 10 s. The NoRH beam width (FWHM) is indicated by the orange ellipse at top-right. Due to low elevation of the Sun during the winter time, the beam size was roughly $17\arcsec \times 30\arcsec$, comparable with the radio source, and therefore, determined an elliptical shape of the latter. 

In Figure \ref{fig1}(c-h) we show intensity images of the sunspot in a sequence of increasing temperatures from six AIA bandpasses: 1700 \AA, 1600 \AA, 304 \AA, 171 \AA, 193 \AA, and 211 \AA. The spatial resolution of data is 0.6 arcsec per pixel. 

Atmospheric Imaging Assembly observes UV continuum in 1700 \AA\ bandpass with the characteristic formation temperature about 5000 K. This bandpass corresponds to the lowest formation layer located close, but still above the temperature-minimum region. According to the sunspot atmospheric model by \cite{Maltby}, the temperature-minimum region is located at about 500 km above $\tau_{500}=1$, corresponding to the upper photosphere/lower chromosphere. The AIA 1600 \AA\ bandpass has a broad response to temperatures ranging from the temperature-minimum to chromospheric line C IV emissions ($4-10\times10^3$ K) in active regions, but the dominant contributor is the UV continuum above the region of temperature-minimum. The 304 \AA\ channel is dominated by the two He II lines with average formation temperature $T=5\times10^4$ K, corresponding to the transition region. EUV light from the Fe IX emission line in the 171 \AA\ bandpass primarily coming from upper transition region/quiet corona material with average formation temperature $T = 6.5\times10^5$ K. The 193 \AA\ channel is expected to be dominated by Fe XII line (average $T=1.5\times10^6$ K) and 211 \AA\ channel is expected to observe Fe XIV line ($T=2\times10^6$ K) for AR observations \citep{O'Dwyer}. 

We did not use the other SDO/AIA bandpasses for the following reasons: visible light continuum observations at 4500 \AA\ are available only with 3600 s time cadence; observations in coronal channels 94 \AA\ and 131 \AA\ had not sufficient quality for this sunspot.

\section{Data analysis}\label{Analysis}
\subsection{Temporal variations of UV/EUV and microwave emission}
It is known that changes in the UV/EUV emission with time must be due to changes in the temperature and density in the emission forming region. Plasma temperature and density variations, in turn, are thought to be evidence some dynamical processes, e.g. a compressible wave passage through the corresponding region. 

Different interpretations for the radio brightness oscillation were proposed, for example, as being caused by the variations in magnetic field strength \citep{Gelfreikh}. \cite{Shibasaki}, however, showed that they can be explained consistently with the EUV observation in terms of the density and temperature fluctuations due to upward-traveling acoustic waves through the third harmonic gyroresonance layer (2000 G for observations at 1.76 cm).

To study temporal variations of UV/EUV emission above the center of the umbra, where the magnetic field is presumably  vertical, we analyzed de-rotated 24-hour series of intensity obtained from one pixel, corresponding to the umbra's center and shown by white cross on the 1700 \AA\ image of Fig. \ref{fig1}(c). The neighbouring pixels show a very similar time variability, hence the data are not affected by the possible variation in pointing. For microwave emission we picked up the value of the maximum brightness temperature, $T_{Bmax}$, located in the center of the elliptical source and, therefore, corresponding to the sunspot umbra center for every synthesized map. 

\begin{figure}
\begin{center}
\includegraphics[totalheight=14cm]{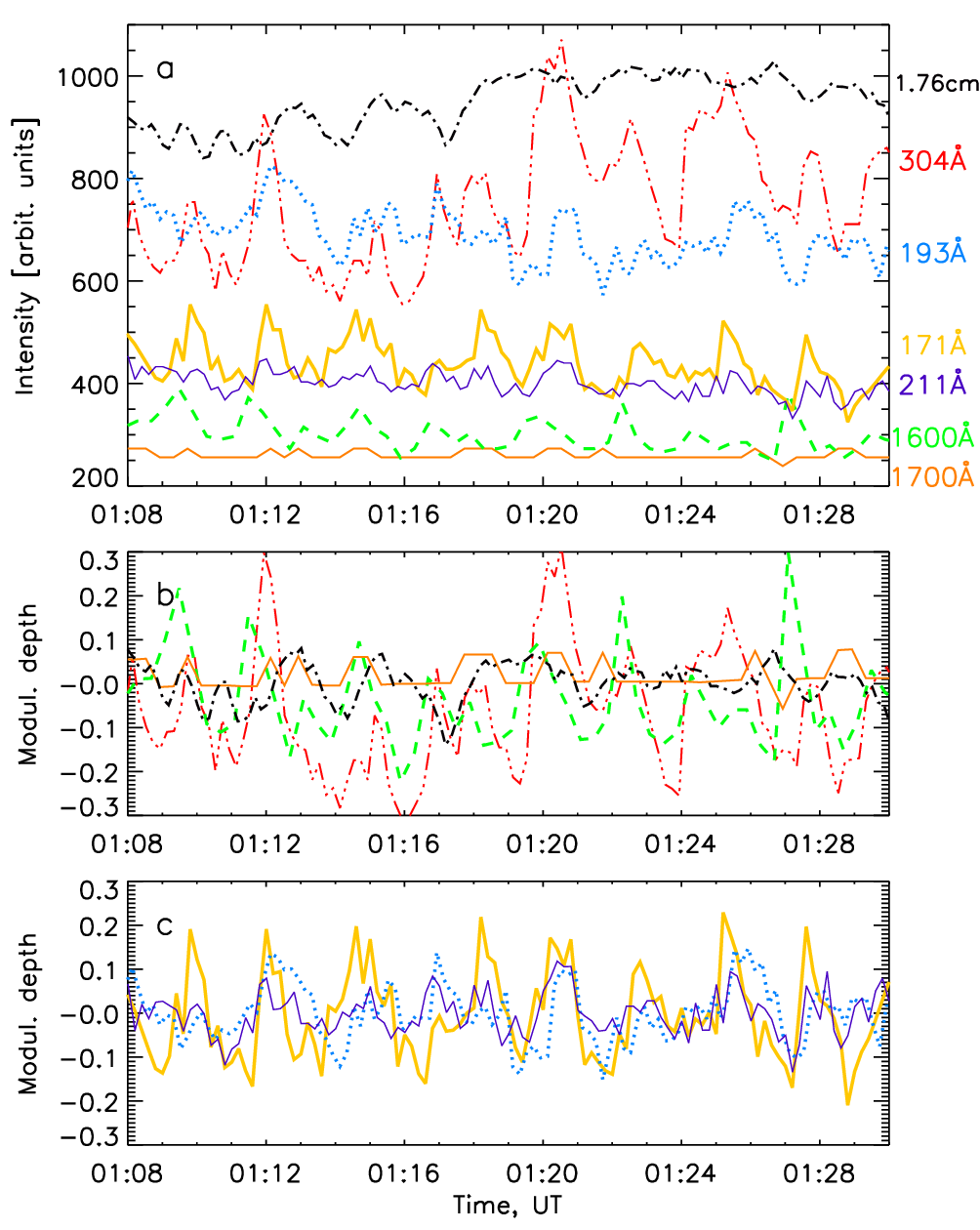}
\end{center}
\caption{(a) The time profiles of intensity above the center of the umbra from six AIA bandpasses and $T_{Bmax}$ at 1.76 cm of 22 min duration. Emission modulation depth at (b) 1700 \AA\ (solid), 1600 \AA\ (dash), 304 \AA\ (dash-dot-dot), and 1.76 cm (dash-dot) and (c) at 171 \AA\ (thick), 193 \AA\ (dots), and 211 \AA\ (thin). For the color version of the figure, different wavelength are shown by colors, corresponding to the key to the right.}
\label{fig2}
\end{figure}

Figure \ref{fig2}(a) shows time profiles, $I(t)$, of the UV/EUV intensity (in arbitrary units) and microwave peak brightness temperature (in K) of 22 minutes duration for clarity. 1600 \AA\ intensity is multiplied by a factor of 3 and NoRH 1.76 cm brightness temperature by a factor ($4\times10^{-3}$). Time cadence is 24 s for 1700 \AA\ and 1600 \AA, 10 s for 1.76 cm and 12 s for other channels. Well-pronounced variations of the emission intensity, with the periods of about two-three minutes, are clearly seen in all channels. 

The time profiles of the modulation depth are shown in Fig. \ref{fig2}(b) for 1700 \AA, 1600 \AA, 304 \AA, and 1.76 cm and in Fig. \ref{fig2}(c) for 171 \AA, 193 \AA, and 211 \AA\ using the same colors as on the top panel. This quantity is calculated as 
\begin{equation}
	\Delta I/I = (I(t)-I_0)/I_0,
\end{equation}
where $I(t)$ is time series of intensity of UV/EUV or microwave brightness temperature. The slowly varying mean signal $I_0$ is obtained by smoothing the signal $I(t)$ over an interval of 14 minutes. The highest modulation depth is found for 304 \AA\ emission, 20\% on the average, and it decreases down to the photosphere, 10\% for 1600 \AA, 8\% for 1700 \AA, and up to the corona with hight, 18\% for 171 \AA, 12\% for 193 \AA, and 8\% for 211\AA.
 
Modulation depth 5\% is found for microwave brightness variation. It corresponds to the change in brightness temperature by $\Delta T_{B}\approx 1.2\times 10^4$ K. These values agree with previous results showing 1 - 5\% of $T_{B}$ variation.

The general trend of the height change in modulation depth of UV/EUV emission also agrees with results of previous studies, which have shown that the oscillation amplitude peaks at transition-region temperatures \citep[e.g.,][]{O'Shea, Marsh}. However, it is rather surprising to find intensity variations with such a high amplitude in coronal channels. Previous studies showed umbral oscillations, observed with TRACE 171 \AA\ and 195 \AA, with much smaller amplitudes (3-8\%) usually guided along the EUV fans of the limited transverse size.  One possibility is the presence in the "coronal" EUV channels of AIA of the emission generated at much lower temperature and hence heights. \cite{Brynildsen} estimated the contributions from the transition region lines to the TRACE 171 \AA\ and 195 \AA\ channel intensity. They found an average value of this contributions equal to 17\% and finally concluded that the significant part of the oscillations in the TRACE 171 \AA\ and 195 \AA\ channels must indeed occur in the corona. \cite{O'Dwyer} performed similar examination for AIA and found that the dominant contribution to three EUV coronal channels under study from the AR plasma comes from the expected ion line formed at the hot, coronal temperatures.

\subsection {Cross-correlation analysis}
The other indirect indicator of the brightness fluctuation origin in each AIA channel can be time shifts between corresponding signals. If the pulsations originate from the same layer of the solar atmosphere we may expect an in-phase behavior for intensity variations. 

\begin{figure}
\begin{center}
\includegraphics[totalheight=7cm]{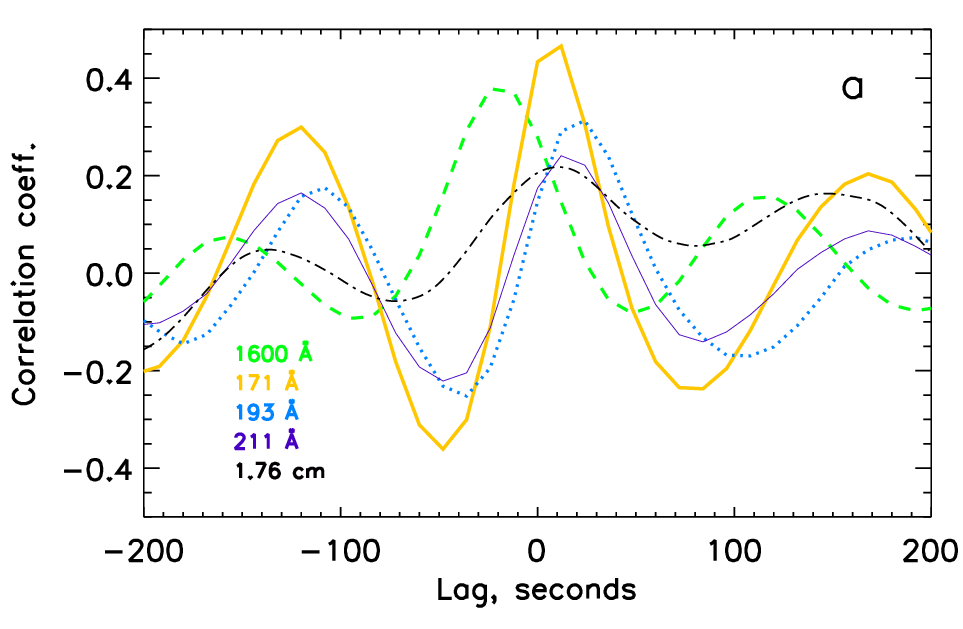}
\includegraphics[totalheight=6cm]{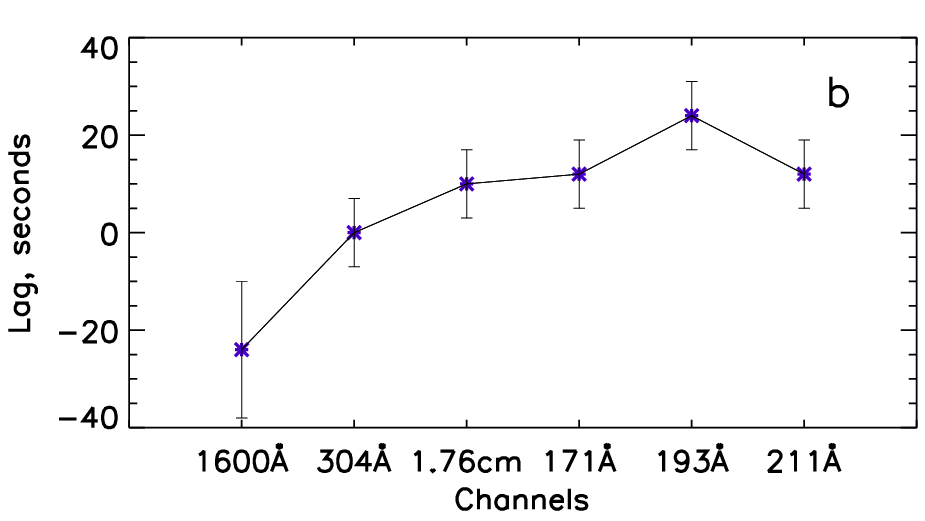}
\end{center}
\caption{(a) Cross correlation coefficients of long time series of pulsations above the umbra's center of each analyzed wavelength with respect to 304 \AA\ as a function of the lag. The same colors and line types as in fig. \ref{fig2} are used to indicate different wavelength. (b) Time lags obtained from cross correlation for each wavelength with the corresponding uncertainties shown by error bars.}
\label{fig3}
\end{figure}

Cross-correlation coefficients as a function of the lag for each analyzed wavelength with respect to 304 \AA\ intensity variation are shown in Figure \ref{fig3}(a) and corresponding delays in Figure \ref{fig3}(b). We used temporal interpolation of the 304 \AA\ time series to match other channel observing times. The precision limit of the obtained delays is estimated by $\sigma=k\times\sigma_s$, where $\sigma_s=\sqrt{\Delta t^2/12}$ is the standard deviation of the sampling error for the uniform time resolution $\Delta t$ and k = 2 for 95\% confidence from the Student-t distribution. These uncertainties $\pm\sigma$ are shown by error bars in Figure \ref{fig3}(b).

Time delays increase with the expected formation height of different channels, except 211 \AA\ channel. Equal delay for 171 \AA\ and 211 \AA\ ($12 s\pm7 s$) supposes that possibly 211 \AA\ intensity has a contribution of cooler temperature lines than expected, or the cooler and hotter plasma structures simply co-exist in the corona at the same height. At the same time, increasing delays found for other wavelength suggest that observed intensity variations in different channels originate from progressively higher altitudes and result from waves propagating upwards along magnetic field lines. The cross-correlation coefficient for 1700 \AA\ bandpass is poor, probably due to very weak emission variation inside the umbra region and nonuniform temporal resolution of downloaded data, and is not shown here. 

If we estimate the expected height separation $\Delta H$ between the two consecutive AIA wavelengths based on the mean formation temperatures mentioned in Section \ref{Obs}, we can use obtained delays to make rough estimates of the average velocity of wave propagation between the two heights. Assuming 1600 \AA\ formation height around temperature-minimum (about 500 km) and 304 \AA\ in the transition region (about 2200 km), $\Delta H_1\approx$ 1700 km gives vertical velocity $v_1\approx$ 70 km s$^{-1}$. This value exceeds the speed of sound in the chromosphere, which lies between 10 and 30 km s$^{-1}$. However, \cite{Abramov-Maximov} found a close value of speed (60 km s$^{-1}$) based on cross-correlation between optical and radio oscillation. 

The height separation $\Delta H_2\approx$ 1000 km between the heights of the 304 \AA\ and 171 \AA\ emission layers gives velocity $v_2\approx$ 83 km s$^{-1}$. This value corresponds to the calculated sound speed with $T=2\times10^5$ K, that is an average temperature between the two formation layers. 

It is rather difficult to define the height of 193 \AA\ emission formation over the sunspot umbra. Bearing in mind the speed of sound (130 km s$^{-1}$) calculated for the 171 \AA\ emission formation temperature ($6.5\times10^5$ K), the delay $12 s\pm7 s$ between the 171 \AA\ and 193 \AA\ emissions gives the minimum value of $\Delta H_3\geq$ 1500 km. Therefore, the 193 \AA\ emission above the umbra corresponds to the corona with the $T\geq1$ MK.
 
Cross-correlation analysis gives $10 s\pm7 s$ of the time delay between the 304 \AA\ and microwave signals, which means that the formation height of radio emission is between AIA 304 \AA\ and 171 \AA\ bandpass. Using $v_2\approx$ 83 km s$^{-1}$ obtained for this height interval, we estimated the altitude of 1.76 cm emission generation, which is approximately 3000 km. The observed value of peak brightness temperature $T_{Bmax}=2.5\times10^5$ K exactly corresponds to the estimated altitude.
 
\begin{figure}
\begin{center}
\includegraphics[totalheight=16cm]{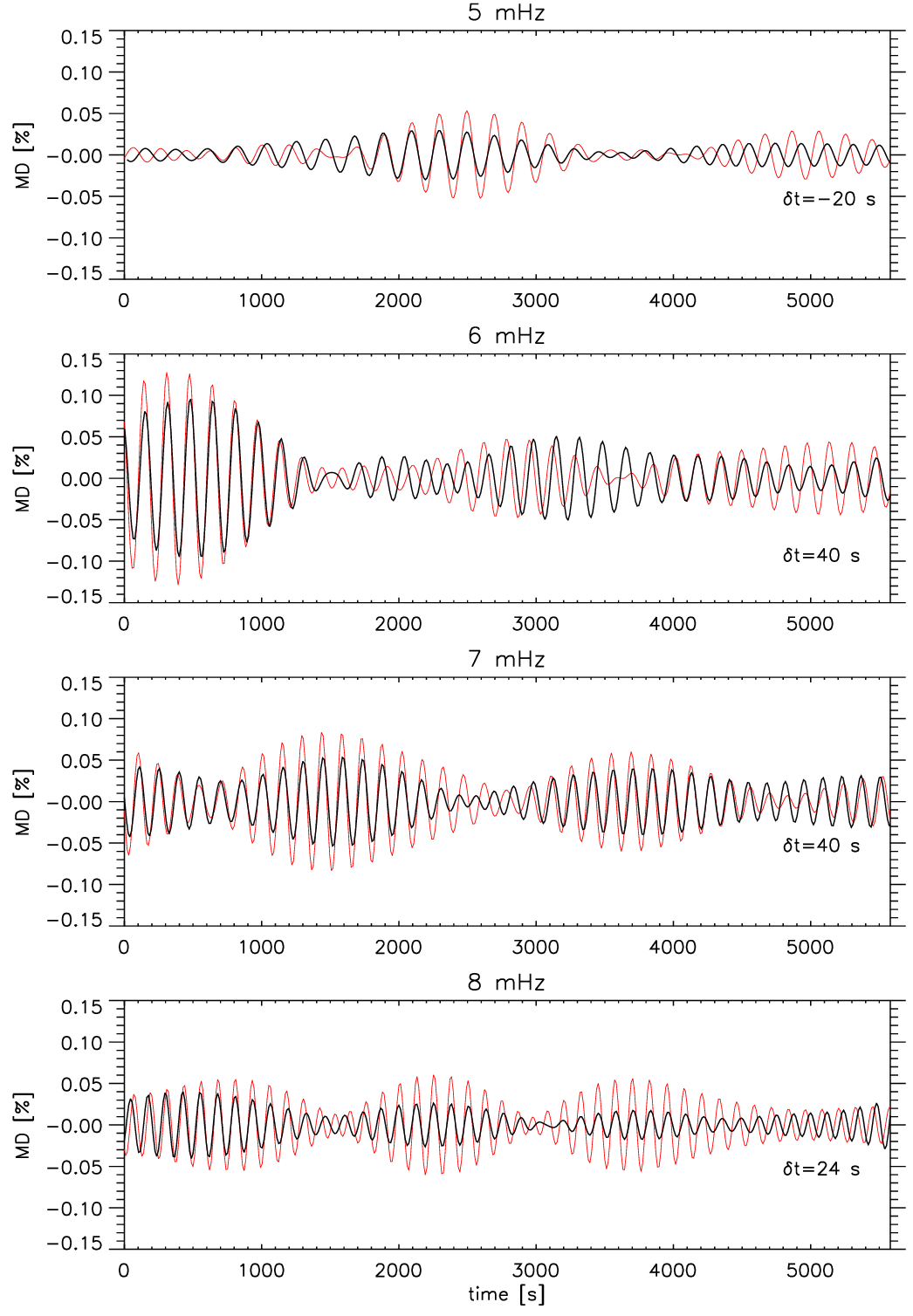}
\end{center}
\caption{Emission modulation depth of signals at 304 \AA\ (red) and 193 \AA\ (black) from the center of the umbra filtered in narrow (1 mHz) frequency bands with the central frequencies at 5 mHz, 6 mHz, 7 mHz and 8 mHz ($\pm$ 0.5 mHz). Signals at two frequencies are overplotted with a time shift $\delta$t to match each other.}
\label{fig4}
\end{figure}

We have to note that the time delays presented in Figure \ref{fig3} are averaged over the frequencies and time. Further analysis reveals that these delays are frequency- and time-dependent. In Figure \ref{fig4} we show the
emission modulation depth of signals at 304 \AA\ (red) and 193 \AA\ (black) at the center of the umbra, shifted in 
time by $\delta$t to match each other. Signals are filtered in narrow (1 mHz) frequency bands with the central 
frequencies at 5 mHz, 6 mHz, 7 mHz and 8 mHz ($\pm$ 0.5 mHz). In this example the 193 \AA\ signal is delayed against 
304 \AA\ signal by 40 s at 6 mHz and 7 mHz, and by 24 s at 8 mHz.  But at 5 mHz we found an inverse delay: the 304 
\AA\ signal is delayed against the 193 \AA\ signal. Moreover, these phase delays are not only frequency-dependent, 
but also time-dependent: though varying the time shift $\delta$t we could adjust the phases of the signals in one wave train, the same phase shift did not do it in the other wave train. The phase shifts were found to vary strongly during the 24-hour observation. Often it was not easy to resolve the $2\pi$ uncertainty.

The fact of the frequency dependence of the delays between the signals measured at different heights agrees with 
previous empirical results obtained from the cross-correlation of photospheric and chromospheric velocity maps 
filtered in narrow bands \citep{Centeno-2006, Centeno-2009}, as well as with the theoretical prediction that 
oscillation modes in dispersive medium propagate with the phase velocity depending on the frequency. The time 
dependence is also consistent with the previous results, e.g. the frequency drifts in the 3-min band \citep{Christopoulou-2003, Sych-2011}. Therefore, detailed analysis of the phase relations between the perturbations at different heightsis of obvious interest. However, it is beyond the scope of this paper and will be addressed in a follow up study.

\subsection{Spectral analysis}\label{Spec}
To study spectral characteristics of intensity variation we used the fast Fourier transform (FFT) method. Because of long intervals of uniform data with high cadence, the spectral analysis was done with high-frequency resolution, 36 $\mu$Hz for NoRH and 12 $\mu$Hz for AIA. Such high spectral resolution in combination with the high spatial and temporal resolution of SDO/AIA allowed us to trace the variations of the spectrum of 3-min oscillations across the umbra. In particular, we were able to trace the so-called low cut-off frequency, which is defined here as a lowest frequency in the 4-9 mHz band for which the Fourier power is three times higher then the noise level.

\begin{figure}
\begin{center}
\includegraphics[totalheight=16cm]{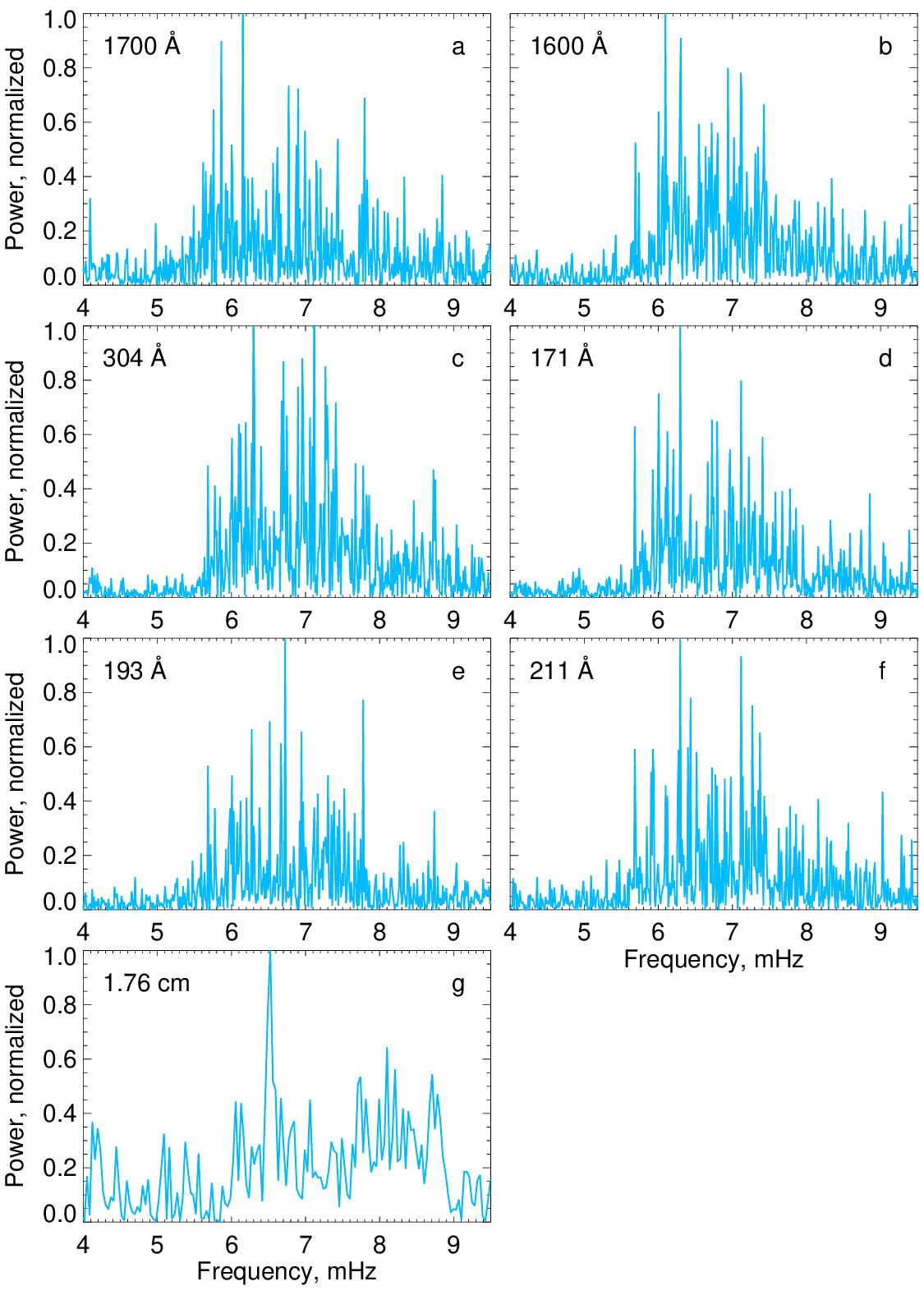}
\end{center}
\caption{Fourier power spectra of the intensity variation above the center of the umbra for all analyzed wavelengths.}
\label{fig5}
\end{figure}

Figure \ref{fig5} represents the power spectra covering 24 hr of UV/EUV data and 7 hr 45 min of microwave data. The power is normalized on its maximum in the band 4-10 mHz. The spectra are shown for the single pixel above the center of the umbra. Most of spectra have dominant frequencies between 6 mHz (period 167 s) and 7 mHz (period 143 s). High-frequency tail extends beyond 8 mHz. All UV/EUV spectra exhibit a sharp increase at 5.7 mHz (equivalent period 175 s), which can be considered as the lower cut-off frequency. 

It is interesting that a similar value (5.8 mHz) was recently found for the maximum cut-off frequency at the axis of a model sunspot in the 3-D nonlinear MHD simulation with the driver located under the photosphere and emitting 5-min waves \citep{Felipe-2010}.

The spectrum of the radio oscillation at 1.76 cm does not show a clear low cut-off. However, the microwave signal has a duration of less than 3 times that the UV and EUV data and comes from much larger area: the difference in the pixel size of AIA and NoRH maps is about 10 times. Nevertheless, similar to all AIA spectra, it has a maximum between 6 and 7 mHz and the increase in the power near 8.8 mHz.

In previous studies the value of the cut-off frequency was often defined as the highest frequency where the phase difference between two signals measured at different heights was zero \citep[e.g.,][]{Centeno-2006, Centeno-2009, Felipe-2011}. As we see from Fig. \ref{fig4}, the phase difference between the 193 \AA\ and 304 \AA\ is positive in the 6-8 mHz band, while is negative at 5 mHz. This confirms that cut-off frequency is located between 5 and 6 mHz. However, the above-mentioned method allows one to determine the cut-off frequency averaged over the umbra, while we are interested in the spatial distribution of this value across the umbra.

\begin{figure}
\begin{center}
\includegraphics[totalheight=10.cm]{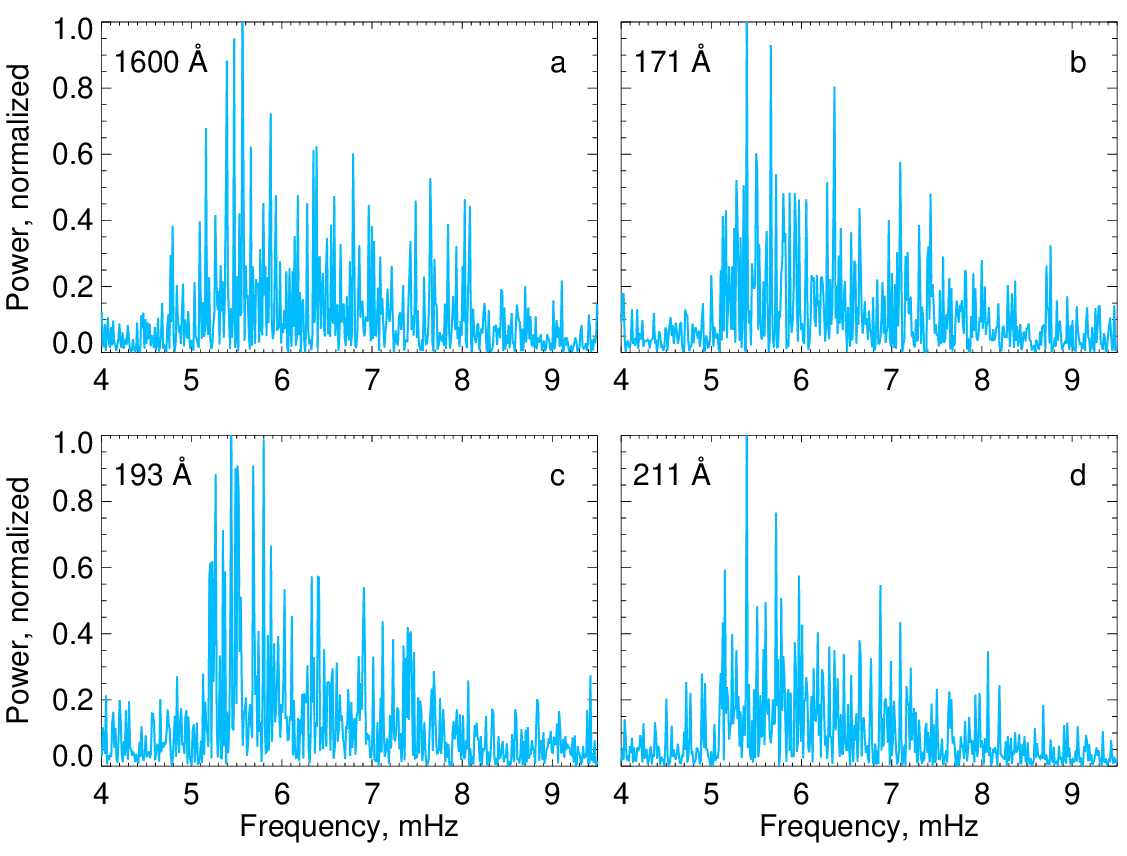}
\end{center}
\caption{Fourier power spectra of the intensity variation near to the umbra/penumbra border from four AIA bandpasses.}
\label{fig6}
\end{figure}

Further analysis has shown that spectral characteristics of oscillations vary across the umbra. Figure \ref{fig6} presents the spectra for pixel located further from the center, near the umbra/penumbra border, it is shown by white asterisk on Fig. \ref{fig1}(c). It is clearly seen that the low frequency cut-off dropped down to 5 mHz and the peak frequency moved down to 5.5 mHz as compared with the umbra's center. The same tendency is found for all other AIA channels.

\begin{figure*}[h!t]
\begin{center}
\includegraphics[totalheight=5.6cm]{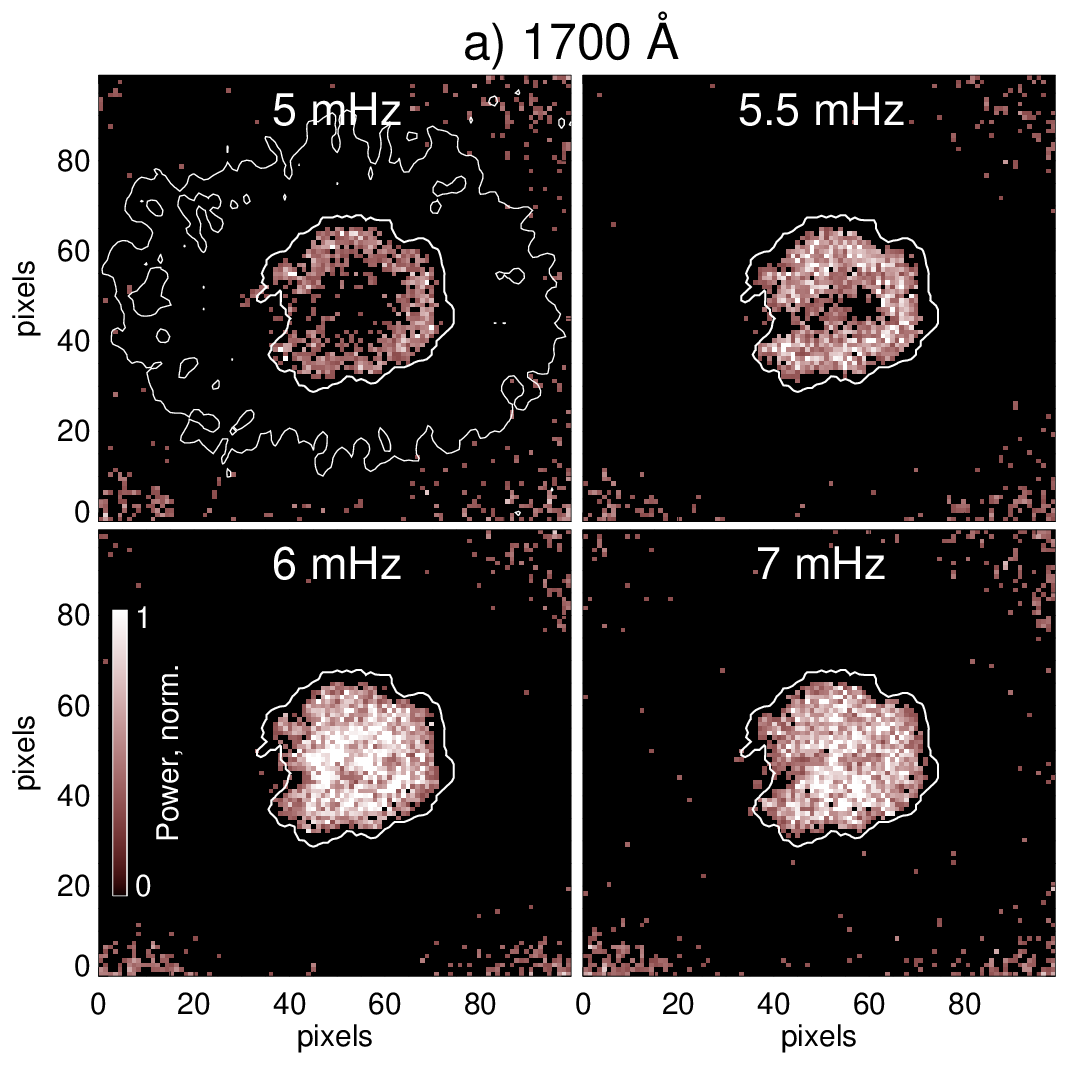}
\includegraphics[totalheight=5.6cm]{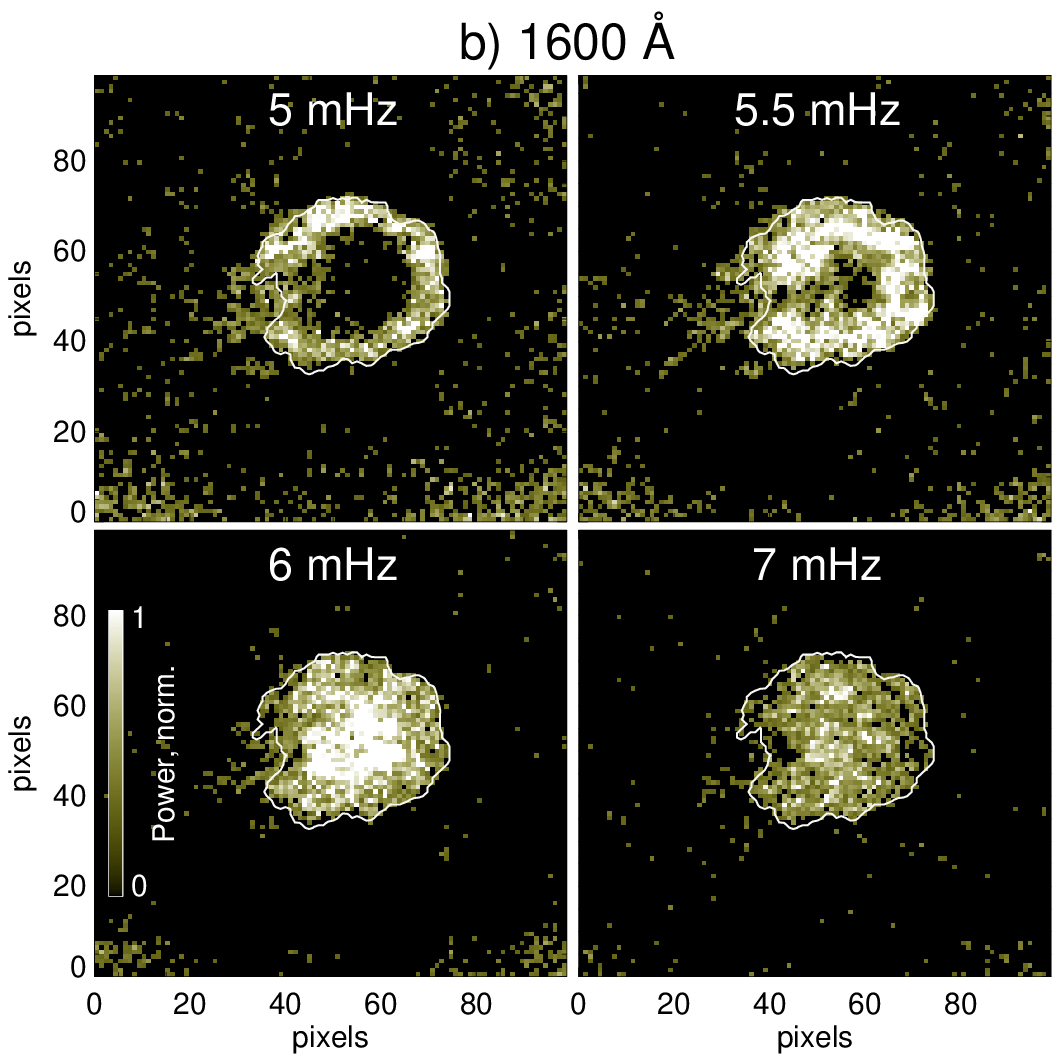}
\includegraphics[totalheight=5.6cm]{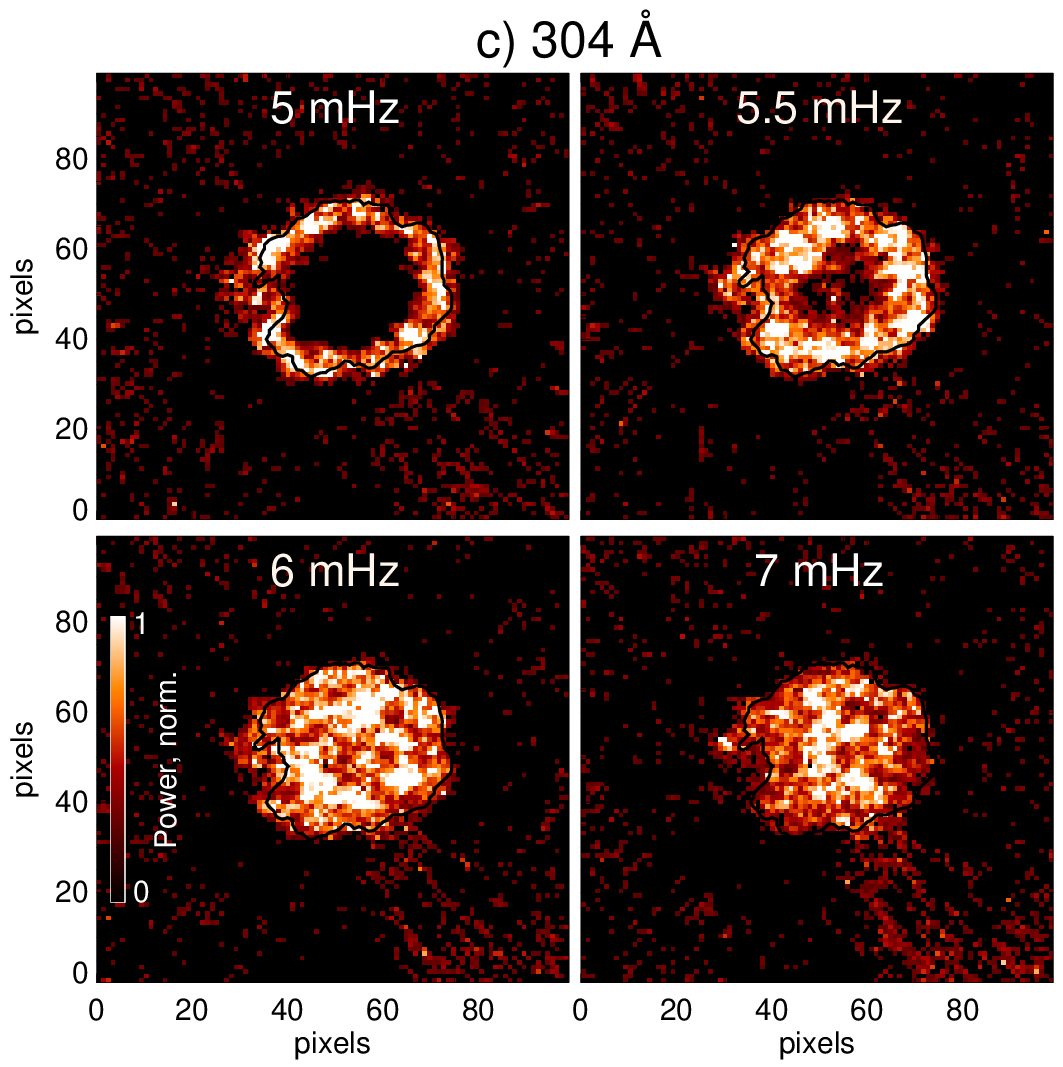}
\includegraphics[totalheight=5.6cm]{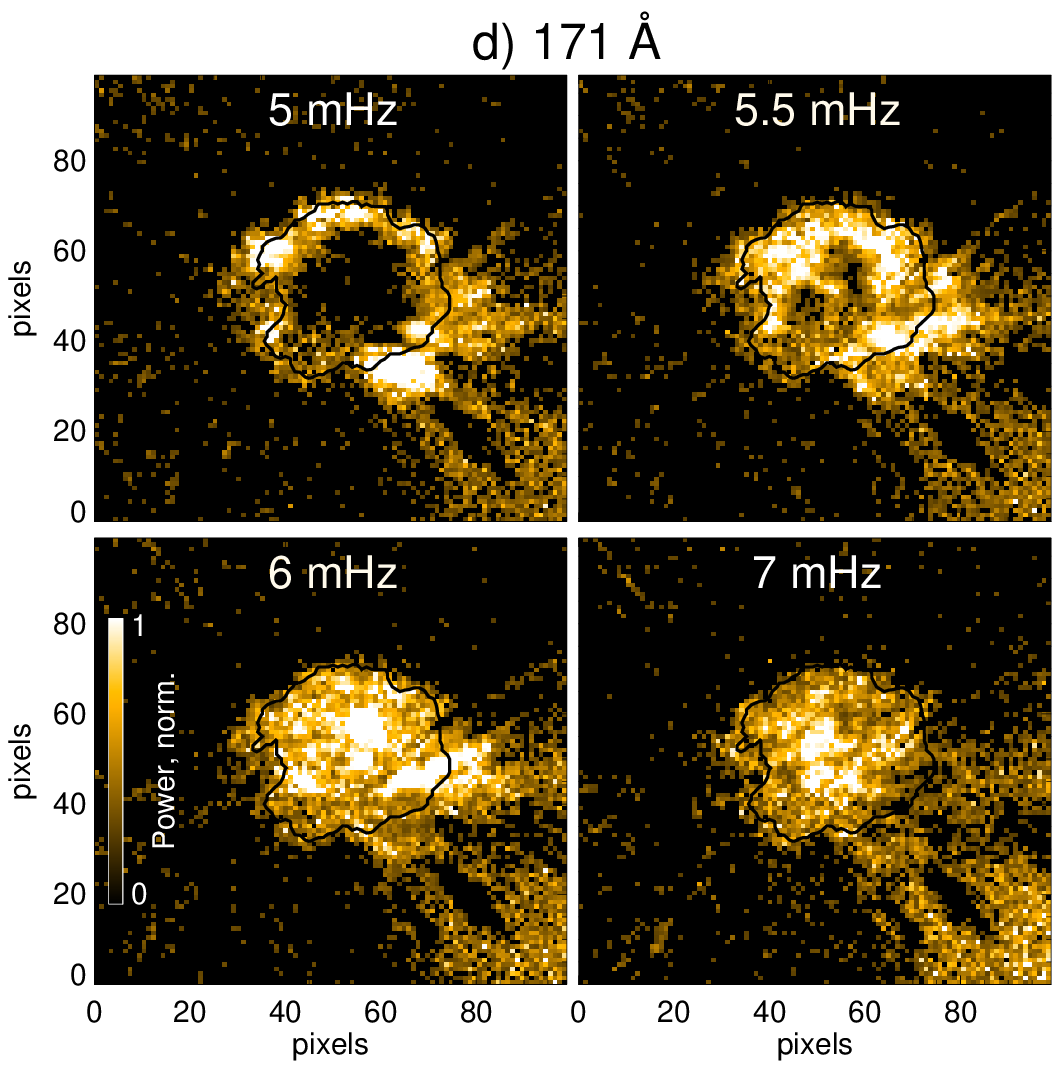}
\includegraphics[totalheight=5.6cm]{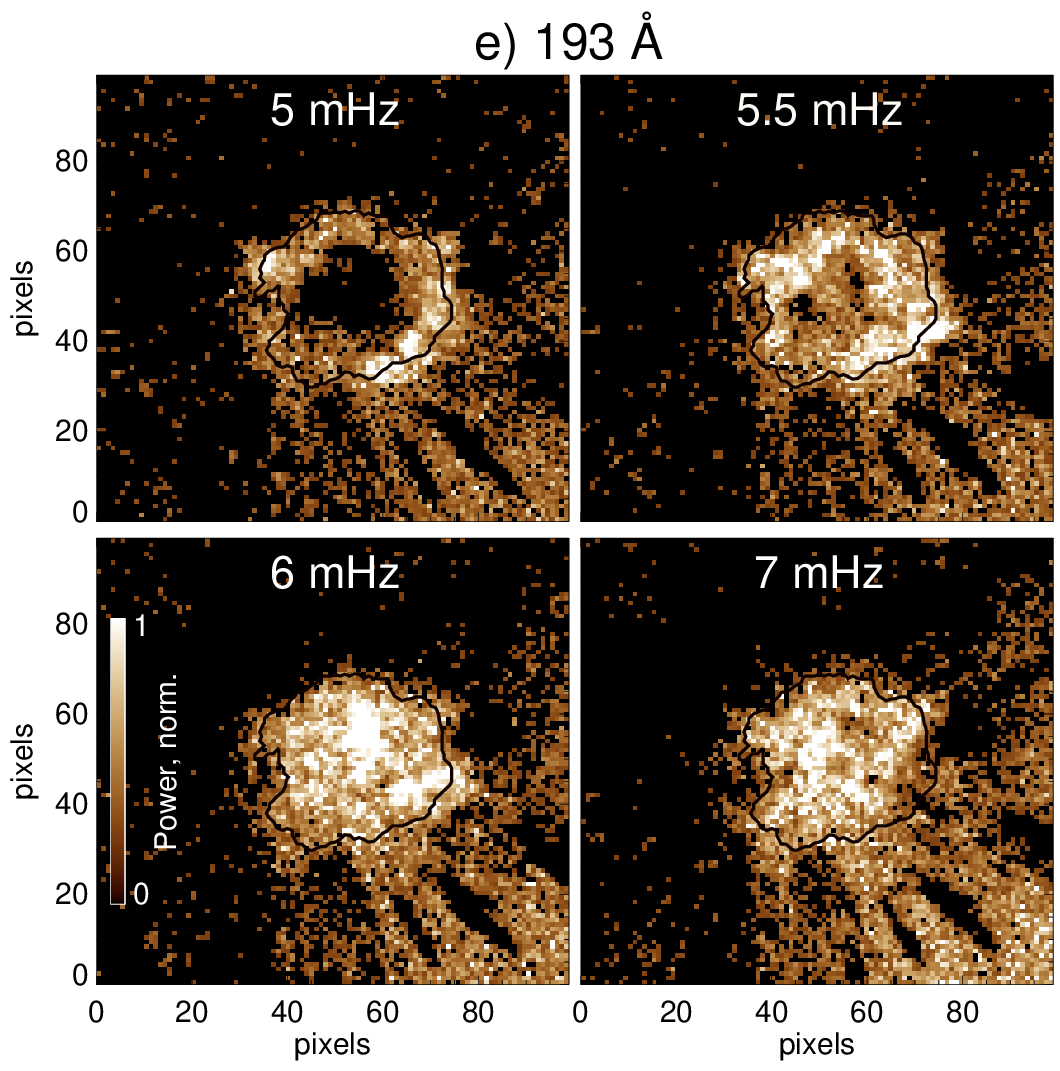}
\includegraphics[totalheight=5.6cm]{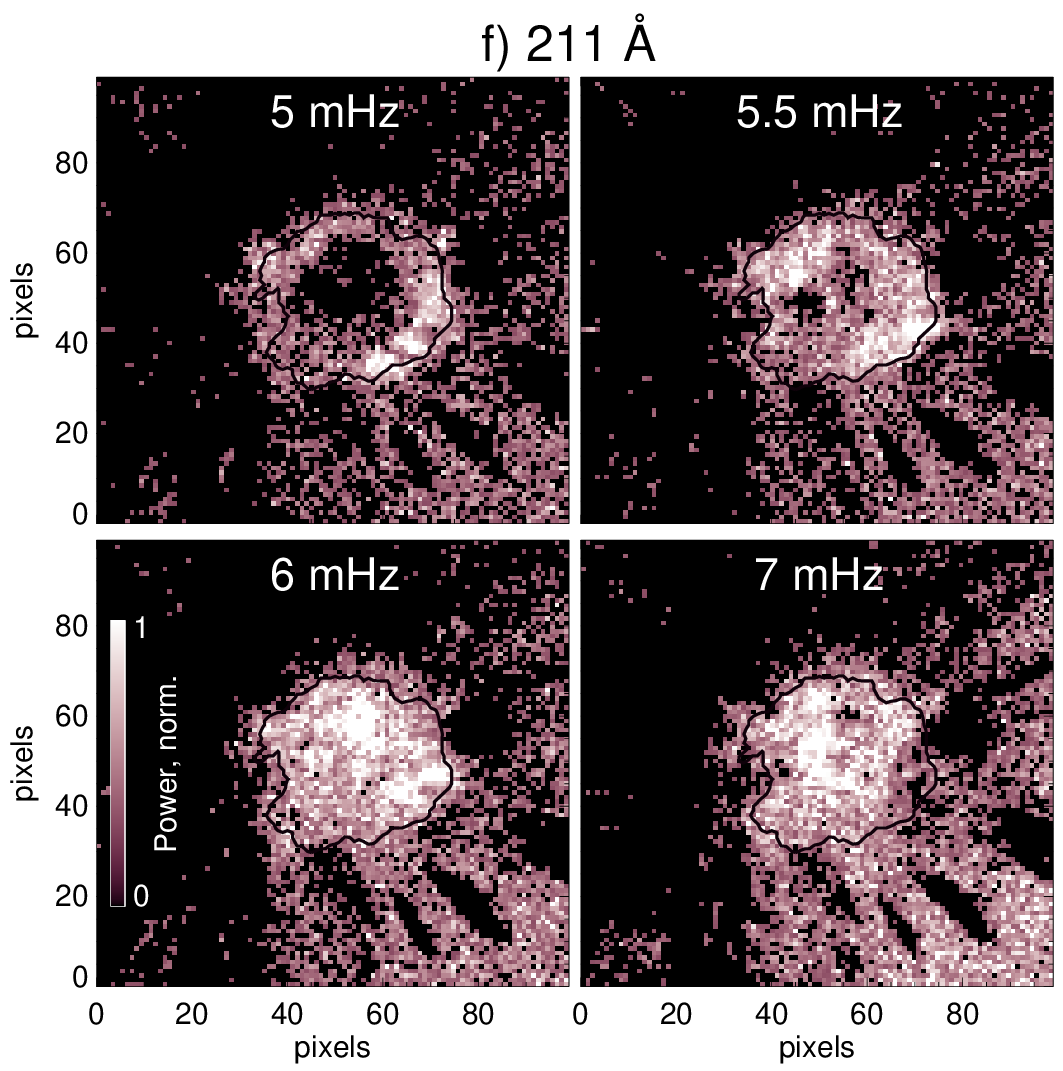}
\end{center}
\caption{Spatial distribution of normalized Fourier power in four frequency bands with the central frequencies at 5 mHz, 5.5 mHz, 6 mHz, and 7 mHz ($\pm 0.2$ mHz in each band). Power grows with brightness. The umbra-penumbra boundary is shown by contour line, white on panels (a-b) and black on panels (c-f).}
\label{fig7}
\end{figure*}

To trace the changes in the spectrum of oscillations for the entire sunspot region we calculated Fourier power spectra for every spatial pixel of the intensity maps. Figure \ref{fig7} gives the spatial distribution of Fourier power in four narrow ($\Delta f=0.4$ mHz) frequency bands with the central frequencies at 5 mHz, 5.5 mHz, 6 mHz, and 7 mHz ($\pm 0.2$ mHz in each band). Individual power spectra from each spatial pixel have been normalized on its spectral maximum. Normalization was performed to aid in the comparison of spectral profiles between spatial regions that exhibit vastly different intensity amplitudes. Only pixels with power larger than three times the noise level are plotted in the power spectrum maps. The umbra-penumbra boundary as observed in 1700 \AA\ continuum is shown in every plot by the contour line, white in panels (a-b) and black in panels (c-f). The external penumbra border is also shown in the upper left plot of panel (a) for reference.

It is clearly seen from Figure \ref{fig7}(a) that Fourier power in all four frequency bands is confined inside the umbra and avoid the penumbral region. In the 5 mHz band it also avoids the central part of the umbra and forms a ring near to the umbra/penumbra border. In the 5.5 band this ring extends inside the umbra, becoming wider. We already saw in Figure \ref{fig5} that oscillations at frequencies $f<5.7$ mHz are absent in the center of the sunspot.
Oscillations in 6 and 7 mHz bands cover the whole umbra, but are more pronounced in its central part. This tendency is seen through all the AIA bandpasses, see Figure \ref{fig7}(b-f). At the same time the oscillating area slightly broadens with the height. The diameter of the oscillating circle is widened by about $\Delta d=1''$ with each subsequent bandpass, expanding beyond the umbra boundary at 304 \AA. In coronal channels 171 \AA\ - 211 \AA\ pulsations with high power outspread to the coronal fan-structures. Higher frequencies, 8 and 9 mHz exhibit even stronger concentration toward the center, but lower power (they are not shown here).

The peculiarities of the spatial distribution of 3-min oscillation frequency found here are consistent with previously obtained results. In particular, with the decrease in the peak frequency from the umbra outwards with the abrupt change in the umbra-penumbra boundary in the chromosphere, reported by \cite{Tziotziou-2006}. Also, the existence of the region with the suppressed power of 3-min oscillations at the center of the umbra has been found by \cite{Jain-2002} in the photospheric Doppler velocity and line-depth data of SOHO/MDI, by
\cite{Nagashima-2007} at lower frequencies (5-5.5 mHz) using Hinode/SOT observations of low-chromospheric Ca II H line intensity, and by \cite{Tziotziou-2007} in their Doppler velocity maps obtained at Ca II 8542 \AA. Here we show that such a region at 5-5.5 mHz exists also at the transition region and coronal heights, and that its size is frequency-dependent.

\section{Interpretation of spectral characteristics}\label{discus}
It is interesting that spectral characteristics of 3-min oscillations, namely, the low cut-off frequency and its variation across the umbra, are similar through all the temperature bandpasses. This finding suggests that the strong magnetic field of a sunspot works as a waveguide for the acoustic waves generated already at the photosphere level and eventually reaching 1 MK corona. 

To understand spectral characteristics of 3-min oscillations above the umbra and tendency of their spatial distribution we will start from the generally accepted idea that these oscillations are caused by magnetoacoustic gravity (MAG) waves, that is the acoustic-like slow mode propagation along magnetic field lines. It has been long known \citep[since][]{Lamb} that due to the presence of gravitational field compressible acoustic waves can propagate upwards through the temperature-minimum only at frequencies above the cut-off frequency:
\begin{equation}
f_{c}=\frac{g\gamma}{2\pi C_{s}},
\label{cut-off}
\end{equation}
where $g$ is gravitational acceleration and
\begin{equation}
C_{s}=\sqrt{\gamma RT/\mu}
\end{equation}
is the speed of sound, T is the temperature, $\gamma$ is the ratio of specific heats, $\mu$ is the mean molecular weight, and R is the gas constant.

Further, \cite{Bel} showed that for MAG waves in the solar atmosphere the cut-off frequency will be affected by the  magnetic field when it is not parallel to the gravitational field. They predicted that in regions of low-$\beta$ plasma ($\beta << 1$), which is the case above the sunspot umbra, the effective gravity on a particular magnetic field line would be decreased by the cosine of the inclination angle $\theta$ of that field line with respect to the direction of gravity (the solar normal), that is 
\begin{equation}
g=g_0 \times cos\theta,
\end{equation}
where $g_{0} = 274$ m s$^{-2}$. Therefore, cut-off frequency 
\begin{equation}
f_{c}\propto cos\theta/\sqrt{T}.
\label{cut-off1}
\end{equation}
Based on the above theoretical predictions one may expect that the change of the observed cut-off frequency across the umbra may result either from variation of temperature in the temperature-minimum layer, $T_{min}$, becoming smaller in the center of the umbra or the magnetic field inclination angle $\theta$ becoming large enough near the umbra/penumbra border. The upper limit of $f_{c}$ is determined by $T_{min}$ on the path of the MAG wave propagating along the vertical field line. 

According to Figure \ref{fig5} the value of cut-off frequency in the center of the umbra, in exactly vertical field, is $f_{c}=5.7$ mHz. If, following \cite{Fleck}, we take $\mu=1.3$ and $\gamma=5/3$ for an ideal neutral gas, using Eq. \ref{cut-off} we obtain $T_{min}\approx3800$ K. The uncertainty of the this estimate is mainly determined by the uncertainty of the mean molecular weight $\mu$: change in $\mu$ by 0.1 gives the change in temperature 300 K. Moreover, the mean molecular weight in turn depends on temperature. 

Near the umbra's boundary observed $f_{c}=5$ mHz. Using proportionality \ref{cut-off1} under the assumption of $T_{min}=const$ gives the inclination angle $\theta\approx30^\circ$ at the peripheral parts of the umbra. This value is consistent with the results of previous study of the radial variation of field inclination on the level of the temperature minimum \citep[e.g. Fig. 3 in][]{McIntosh}.

On the other hand, it is reasonable to assume that the minimum temperature grows from the middle to the peripheral parts of the umbra. Under the assumption of vertical field through all the umbra the observed value $f_{c}=5$ mHz gives $T_{min}\approx4900$ K. This value is too large even for the quiet-Sun atmosphere \citep{Avrett}. Therefore, we think that the temperature increase has a smaller effect on the wave frequency distribution across the umbra than the magnetic field inclination, though both effects modify the cut-off frequency in a similar way. 

The gradual broadening of the area occupied by the oscillations with the height of the solar atmosphere agrees well with the above interpretation. Roughly speaking, those field lines inclined 30$^\circ$ to the solar normal must form a wider circle in the transition region than at the temperature-minimum level. An estimate using the simplifying assumption that the magnetic field remains linear between the two formation heights (there is no field curvature), gives 1-2$''$ of the circle expansion. This estimate well agrees with the observed value $\Delta d$ shown in Section \ref{Spec}. 
 
Observational demonstration of the acoustic cut-off frequency modification in the penumbra area using TRACE observations presented by \cite{McIntosh}. Results obtained in this work can be considered as a demonstration of the same effect in the umbral area. Nevertheless, magnetic field extrapolation would be useful to confirm it. 

The other possible explanation of ring-like frequency-spatial structure can be done in the frame of the chromospheric acoustic resonator model \citep{Zhugzhda, botha}. Indeed, the increase in the angle of the guiding magnetic field from vertical, the size of the resonator grows and hence the resonant frequency decreases.

\section{Conclusion}\label{concl}
Three-min oscillations above a sunspot have been studied both at short (UV/EUV) and long (microwave) wavelengths. 
Intensity variations with $f=5-9$ mHz are found in all analyzed UV/EUV bandpasses as well as in microwaves (1.76 cm) with the similar spectral characteristics. The largest modulation depth (20\%) is found in the transition region (304 \AA\ channel) and it decreases down to the photosphere and up to the corona. Time delays between oscillations at different wavelength above the umbra's center increase with the expected formation height of the corresponding emission (except for 211 \AA). This confirms that observed intensity variations of both UV/EUV and microwave emission result from waves propagating upwards. Measured delays are used to estimate average velocities of wave propogation at different layer of solar atmosphere and the altitude of microwave emission generation.

High spatial, temporal and spectral resolution of SDO/AIA allowed us to trace the variations of the cut-off frequency and spectrum of oscillations across the umbra. We found that in the range 5-9 mHz high-frequency oscillations are more pronounced closer to the umbra's center, while the low-frequencies concentrate to the peripheral parts. Accordingly, the cut-off frequency decreases from 5.7 mHz to 5.0 mHz. We interpreted this discovery as a manifestation of variation of the magnetic field inclination across the umbra at the level of temperature-minimum. This interpretation has been used for diagnostics of sunspot atmosphere on this level.

The similar spectral characteristics of 3-min oscillations were found through all the temperature bandpasses. This suggests that the strong magnetic field of a sunspot works as a waveguide for the acoustic waves propagating from the photosphere level and eventually reaching 1 MK corona. 

\acknowledgements
{Author is grateful to Drs. M. Shimojo and  W.R. Ashworth for useful discussions. We are thankful to anonymous referee for suggesting improvements to an earlier versions of this paper. The work was partly supported by RFBR grants No.10-02-00153-a, 11-02-91175, and 11-02-10000-k. The research carried out by Dr. Robert Sych at NAOC was supported by the Chinese Academy of Sciences Visiting Professorship for Senior International Scientists, grant  No. 2010T2J24.
}

\end{document}